# Analytical calculation of the minimum wind speed required to sustain wind-blown sand on Earth and Mars

Jasper F. Kok[1]

[1]Advanced Study Program, National Center for Atmospheric Research, Boulder, Colorado, USA

**ABSTRACT**

The wind-driven hopping motion of sand grains, known as saltation, forms dunes and ripples and ejects fine dust particles into the atmosphere on both Earth and Mars. While the wind speed at which saltation is initiated, the 'fluid threshold,' has been studied extensively, the wind speed at which it is halted, the 'impact threshold,' has been poorly quantified for Mars conditions. We present an analytical model of the impact threshold, which we show to be in agreement with measurements and recent numerical simulations for Earth conditions. For Mars conditions, we find that the impact threshold is approximately an order of magnitude below the fluid threshold, in agreement with previous studies. Saltation on Mars can thus be sustained at wind speeds an order of magnitude less than required to initiate it, leading to the occurrence of 'hysteresis.' These results confirm earlier simulations with a detailed numerical saltation model, and have important implications for the formation of sand dunes, ripples, and dust storms on Mars.

# ARTICLE

## 1. Introduction

The wind-driven hopping motion of sand grains, which occurs in arid areas on both Earth and Mars, is known as saltation (Figure 1). Saltation is the driving force behind the formation of dunes on both planets [*Bagnold*, 1941], is an important agent of wind erosion [*Greeley and Iversen*, 1985], and ejects fine dust particles (aerosols) into the atmosphere, which greatly affects weather and climate on both planets [*Goudie and Middleton*, 2006; *Zurek et al.*, 1992]. Finally, electric fields generated in saltation and dust storms on Mars [*Kok and Renno*, 2009a] could affect atmospheric chemistry [*Atreya et al.*, 2006] and produce electric discharges [*Ruf et al.*, 2009].

Saltation is initiated when the wind shear velocity (a measure of the wind stress $\tau$ defined as $u^* = \sqrt{\tau/\rho_a}$, where $\rho_a$ denotes the air density) exceeds the 'fluid threshold' at which surface particles are lifted [*Bagnold*, 1941]. Once initiated, saltation is maintained by the splashing (Fig. 1) of surface particles by impacting saltating particles [*Anderson and Haff*, 1991; *Kok and Renno*, 2009b]. This process transfers momentum to the soil bed more efficiently than fluid drag does, and thus allows saltation to be sustained at values of the wind shear velocity below the fluid threshold. The lowest shear velocity at which saltation can be sustained in this manner is termed the 'impact threshold' [*Bagnold*, 1941]. For loose sand on Earth, the ratio of the impact and fluid thresholds is approximately 0.82 [*Bagnold*, 1937]. No measurements of the impact threshold for Mars conditions exist, but several recent studies indicate that the ratio of the impact to fluid thresholds is substantially lower than on Earth [*Claudin and Andreotti*, 2006; *Almeida et al.*, 2008; *Kok*, 2010].

Here, we use an analytical model to derive the impact threshold on both Earth and Mars. The results for Earth are in agreement with measurements and the results of a recent numerical study [*Kok and Renno*, 2009b]. For Mars conditions, we find that the impact threshold is approximately an order of magnitude below the Martian fluid threshold, and consistent with the recent numerical results of *Kok* [2010].

The small ratio of the impact and fluid thresholds on Mars allows saltation transport there to occur well below the fluid threshold, which has important implications for the formation of dunes and ripples and possibly also for the emission of dust aerosols [*Kok*, 2010]. We derive an approximate expression of the Martian impact threshold for use in future studies.

## 2. Analytical model of the impact threshold

We start our analytical derivation of the impact threshold by considering the typical trajectory of a particle of size $D_p$ saltating over a bed of similar particles. The trajectory of the particle is determined mainly by gravity and fluid drag, and we thus neglect secondary forces due to particle spin [*White and Schulz*, 1977], electrostatics [*Kok and Renno*, 2008; *Kok and Lacks*, 2009], turbulence [*Kok and Renno*, 2009b], and interparticle collisions [*Huang et al.*, 2007]. In steady-state saltation, the particle concentration stays constant, such that this typical trajectory must yield an average impact speed that is as likely to result in the loss of the saltating particle to the soil bed, as it is to splash up a second particle [*Kok and Renno*, 2009b]. Below, we thus calculate the value of the impact threshold on Earth and Mars from the condition that the speed gained by the particle from the wind profile at the impact threshold results in a constant particle concentration.

The horizontal speed gained by the particle over its trajectory can be expressed as

$$\overline{\Delta v_x} = \overline{v_{imp}} - \overline{v_{x0}} = \frac{\overline{\overline{F_d}}\,\overline{t_{hop}}}{m}, \quad (1)$$

where $\overline{v_{imp}}$ and $\overline{v_{x0}}$ are the impact speed and horizontal lift-off speed averaged over the ensemble of saltating particles, $\overline{t_{hop}}$ is the ensemble-averaged hop duration, and $m$ is the particle mass. $\overline{\overline{F_d}}$ is the average drag force in the horizontal direction, which has a double overbar to represent the average over the trajectory of the 'average' saltating particle. Note that in deriving Eq. (1), we have neglected the vertical component of the impact speed, which is much smaller than the horizontal component since particles strike the soil at angles of only ~5 – 15° from horizontal [*Rice et al.*, 1995; *Wang et al.*, 2008].

The trajectory-averaged drag force $\overline{\overline{F_d}}$ can be approximated as

$$\overline{\overline{F_d}} \approx \frac{\pi D_p^2}{8} \rho_a \overline{\overline{C_D}} \left(\overline{\overline{U_x}} - \overline{\overline{v_x}}\right) \overline{\overline{v_R}}, \quad (2)$$

where $\overline{\overline{v_x}}$ and $\overline{\overline{U_x}}$ are respectively the trajectory-averaged horizontal particle speed and wind speed, for which we derive approximate expressions below. Since the vertical particle speed is generally much smaller than the difference between the wind speed and the horizontal particle speed [*Bagnold*, 1941], a reasonable approximation of the average relative speed $\overline{\overline{v_R}}$ between the particle and the wind is

$$\overline{\overline{v_R}} = \left|\overline{\overline{U_x}} - \overline{\overline{v_x}}\right|. \quad (3)$$

The average drag coefficient $\overline{\overline{C_D}}$ for irregularly-shaped sand particles is approximately given by [*Cheng*, 1997; *Kok and Renno*, 2009b]

$$\overline{\overline{C_D}} = \left[\left(\frac{32}{\overline{\overline{Re}}}\right)^{2/3} + 1\right]^{3/2}, \quad (4)$$

where $\overline{\overline{Re}}$ is the average particle Reynolds number,

$$\overline{\overline{Re}} = \frac{\rho_a D_p \overline{\overline{v_R}}}{\mu}, \quad (5)$$

with the viscosity $\mu$ given by the Sutherland relation

$$\mu = \mu_0 \left[\frac{T_0 + C}{T + C}\right](T/T_0)^{3/2}. \quad (6)$$

For $CO_2$ we have $\mu_0 = 1.48 \times 10^{-5}$ kg/m·s, $C = 240$, and $T_0 = 293.15$ K [*Crane Company*, 1988]. Finally, the average horizontal wind speed $\overline{\overline{U_x}}$ depends on the wind profile at the impact threshold. Conveniently, this wind profile is unperturbed from that in the absence of saltation, since the particle concentration approaches zero near the impact threshold. The average wind speed experienced by a saltating particle over its trajectory can thus be approximated from the logarithmic "law of the wall," which describes the turbulent flow over an aerodynamically rough surface [*Prandtl*, 1935]:

$$\overline{\overline{U_x}} = \frac{u*_{it}}{\kappa} \ln\left(\frac{\overline{\overline{z}}}{z_0}\right), \quad (7)$$

where $u*_{it}$ is the impact threshold, $\kappa = 0.40$ is the von Kármán constant, $\overline{\overline{z}}$ is the average height of the particle above the surface during its hop, and $z_0 = D_p/30$ is the aerodynamic roughness length of a flat bed of sand particles [*Nikuradse*, 1933]. Inserting Eqs. (2-7) into Eq. (1) and solving for the impact threshold then yields

$$u*_{it} = \frac{\kappa}{\ln(\overline{\overline{z}}/z_0)}\left[\overline{\overline{v_x}} + \sqrt{\frac{\overline{v_{imp}} - \overline{v_{x0}}}{C\overline{t_{hop}}}}\right], \quad (8)$$

where

$$C = \frac{3\rho_a \overline{\overline{C_D}}}{4\rho_p D_p}. \quad (9)$$

The impact threshold thus depends on the average horizontal lift-off ($\overline{v_{x0}}$) and impact ($\overline{v_{imp}}$) speeds, the duration of an 'average' saltation hop ($\overline{t_{hop}}$), and the mean height ($\overline{\overline{z}}$) and horizontal particle speed ($\overline{\overline{v_x}}$) of the 'average' saltation trajectory. We derive approximate expressions for these quantities in the subsequent sections.

## 2.1 Average particle speed at impact and lift-off.

We approximate the average particle speed at impact and lift-off by using the steady-state requirement that the number of splashed particles ($N_{spl}$) must balance the number of saltating particles lost to the soil bed ($N_{loss}$) [*Kok and Renno*, 2009b]. These quantities can respectively be approximated by [*Anderson and Haff*, 1991; *Kok and Renno*, 2009b]

$$N_{spl} = \sum_i \frac{a}{\sqrt{gD_p}} v^i_{imp}, \quad (10a)$$

$$N_{loss} = \sum_i 1 - P_{reb}(v^i_{imp}), \quad (10b)$$

where $i$ sums over all saltating particles that impact the soil surface per unit time and unit area, and $g$ is the gravitational acceleration. The function $P_{reb}$ describes the chance that a saltating particle will rebound, which we approximate by [*Anderson and Haff*, 1991]

$$P_{reb} = B[1 - \exp(-\gamma v_{imp})]. \quad (11)$$

The parameter values $a = 0.020$, $B = 0.96$, and $\gamma = 1$ s/m were determined by the recent numerical modeling study of *Kok and Renno* [2009b] and are consistent with laboratory and numerical experiments [e.g., *Anderson and Haff*, 1991; *Rice et al.*, 1995]. Note that Eqs. (10, 11) neglect cohesive forces and thus overestimate $N_{spl}$ for particles smaller than ~100 – 150 μm, which can experience cohesive forces that exceed the gravitational force [*Greeley and Iversen*, 1985; *Shao and Lu*, 2000; *Kok and Renno*, 2006].

In steady-state saltation we have that $N_{spl} = N_{loss}$, and thus

$$\sum_i \frac{a}{\sqrt{gD_p}} v^i_{imp} = \sum_i 1 - B\left[1 - \exp\left(-\gamma v^i_{imp}\right)\right]. \tag{12}$$

We solve Eq. (12) for the ensemble-averaged impact speed $\overline{v_{imp}}$ by assuming a plausible distribution of the impact speed $v_{imp}$. Simulations with our recent comprehensive numerical saltation model, called COMSALT [*Kok and Renno*, 2009b], show that the impact speed is approximately exponentially distributed (see Fig. 2), and we thus assume:

$$\Omega(v_{imp}) = \frac{\exp(-v_{imp}/\overline{v_{imp}})}{\overline{v_{imp}}}, \tag{13}$$

where $\Omega(v_{imp})$ is the probability density function of the impact speed. Note that assuming different plausible impact speed distributions only slightly changes the results presented below. Replacing the sum in Eq. (12) with a continuous integral scaled by the probability distribution function of Eq. (13), and solving for the average impact speed then yields

$$\overline{v_{imp}} = \frac{(1-B)\sqrt{gD_p}}{2a} - \frac{1}{2\gamma} + \sqrt{\frac{1}{4\gamma^2} + \left(\frac{1-B}{2a}\right)^2 gD_p + \frac{(1+B)\sqrt{gD_p}}{2a\gamma}}. \tag{14}$$

Note that Eq. (14) indicates that the average impact speed stays constant with shear velocity [*Ungar and Haff*, 1987; *Kok and Renno*, 2009b], which was indeed confirmed by recent wind-tunnel measurements [*Rasmussen and Sorensen*, 2008; *Creyssels et al.*, 2009]. Moreover, the average impact speed calculated from Eq. (14) is ~1.2 m/s for 250 μm particles on Earth, which is consistent with these wind-tunnel measurements [*Rasmussen and Sorensen*, 2008; *Creyssels et al.*, 2009; *Kok*, 2010].

In order to quantify the total amount of horizontal momentum gained by the particle through wind drag (see Eq. 1), we also need to approximate the average lift-off speed. Particles can lift-off from the surface either as splash ejecta or as rebounds following a surface collision. The average lift-off speed is thus the sum of two components,

$$\overline{v_{lift}} = \int_0^\infty P_{reb}(v_{imp})\Omega(v_{imp})\overline{\alpha_R}v_{imp}dv_{imp} + f_{ej}\overline{v_{ej}}, \tag{15}$$

where $\overline{\alpha_R} \approx 0.55$ [*Rice et al.*, 1995] is the average restitution coefficient (i.e., the fraction of momentum retained by a saltating particle upon colliding with the soil surface), and $f_{ej}$ is the fraction of the total number of particles lifting off from the surface that are splash ejecta. We obtain this fraction by combining Eqs. (10a) and (13) and using that $N_{loss} = N_{spl}$, which yields

$$f_{ej} = \frac{a}{\sqrt{gD}} \overline{v_{imp}}. \tag{16}$$

Furthermore, the average ejected particle speed $\overline{v_{ej}}$ was derived in Section 2.2 of *Kok and Renno* [2009b] and approximately equals

$$\overline{v_{ej}} = \frac{\overline{\alpha_{ej}}\sqrt{gD}}{a}\left[1 - \exp\left(-\frac{\overline{v_{imp}}}{40\sqrt{gD}}\right)\right], \tag{17}$$

where $\overline{\alpha_{ej}} \approx 0.15$ is the average fraction of impacting momentum that is contained in splash ejecta [*Rice et al.*, 1995; *Kok and Renno*, 2009b]. Substituting Eq. (17) into Eq. (15) and evaluating the integral then yields the average lift-off speed as

$$\overline{v_{\text{lift}}} = B\overline{\alpha_R}\,\overline{v_{\text{imp}}}\left[1 - \frac{1}{(1+\gamma\overline{v_{\text{imp}}})^2}\right] + \frac{a\overline{v_{\text{imp}}}}{\sqrt{gD}}\,\overline{v_{\text{ej}}}. \tag{18}$$

We then finally obtain the average horizontal and vertical lift-off speeds as

$$\overline{v_{x0}} = \overline{v_{\text{lift}}}\cos\overline{\theta_{\text{lift}}}, \tag{19a}$$

$$\overline{v_{z0}} = \overline{v_{\text{lift}}}\sin\overline{\theta_{\text{lift}}}, \tag{19b}$$

where $\overline{\theta_{\text{lift}}} \approx 40°$ is the average lift-off angle [*Anderson and Haff*, 1991; *Rice et al.*, 1995; *Kok and Renno*, 2009b].

## 2.2 The average hop time and particle height

In addition to the above expressions for the average impact and lift-off speed, we also require expressions for the average hop time and particle height. We derive these from the equation of motion in the vertical direction,

$$\frac{dv_z}{dt} = -\frac{v_z}{\tau_r} - g, \tag{20}$$

where the relaxation time $\tau_r = (C\overline{\overline{v_R}})^{-1}$ is a measure of how quickly the particle speed approaches the fluid speed. The solution to Eq. (20) is

$$v_z = [\overline{v_{z0}} + g\tau_r]\exp(-t/\tau_r) - g\tau_r. \tag{21}$$

We now obtain the average hop duration $\overline{t_{\text{hop}}}$ by integrating Eq. (21) to obtain $z$ and solving for the time where $z = 0$, which yields the recursive formula

$$\overline{t_{\text{hop}}} = \left[\frac{\overline{v_{z0}}}{g} + \tau_r\right]\left[1 - \exp(-\overline{t_{\text{hop}}}/\tau_r)\right]. \tag{22}$$

Additionally, we approximate the average particle height as $\overline{\overline{z}} = \overline{z_{\text{max}}}/2$, where we obtain the maximum trajectory height $\overline{z_{\text{max}}}$ by solving for the time $t$ where $v_z = 0$ in Eq. (21) and substituting that into the expression for $z$, which yields

$$\overline{z_{\text{max}}} = \overline{v_{z0}}\tau_r + g\tau_r^2 \ln\left(\frac{g\tau_r}{\overline{v_{z0}} + g\tau_r}\right). \tag{23}$$

## 2.3 The average horizontal particle speed

The last ingredient we need in order to use Eq. (8) to calculate the impact threshold is an expression for the average horizontal particle speed $\overline{\overline{v_x}}$. We derive this from the horizontal equation of motion (see Eqs. 2 and 3),

$$\frac{dv_x}{dt} = -C(v_x - \overline{\overline{U_x}})|v_x - \overline{\overline{U_x}}|, \tag{24}$$

solving for $v_x$ by using that $\overline{\overline{U_x}} > v_x$ yields

$$\overline{\overline{v_x}} = \frac{Ct\overline{\overline{U_x}}(\overline{\overline{U_x}} - \overline{v_{x0}}) + \overline{v_{x0}}}{1 + Ct(\overline{\overline{U_x}} - \overline{v_{x0}})}. \tag{25}$$

We now average Eq. (25) over the hop duration (Eq. 22) to obtain the average horizontal particle speed,

$$\overline{\overline{v_x}} = \overline{\overline{U_x}} - \frac{1}{C\overline{t_{hop}}} \ln\left[1 + C\overline{t_{hop}}\left(\overline{\overline{U_x}} - \overline{v_{x0}}\right)\right].$$  (26)

### 3. Results

With the expressions for the average impact speed (Eq. 14), lift-off speed (Eq. 19a), horizontal particle speed (Eq. 26), hop time (Eq. 22), and particle height (Eq. 23) in hand, we now obtain the impact threshold by iteratively solving Eqs. (2-5, 7-9, 14, 18, 19, 22, 23, 26). The solution of this procedure is plotted in Fig. 3a for Earth ambient conditions (i.e., $g = 9.81$ m/s$^2$, $P = 101325$ Pa, $T = 300$ K, and $\rho_p = 2650$ kg/m$^3$) and is compared with measurements of the impact threshold [*Bagnold*, 1937; *Iversen and Rasmussen*, 1994] and simulations with the numerical saltation model COMSALT [*Kok and Renno*, 2009b]. The analytical model approximately reproduces the measurements, although the averaging procedures it employs result in a somewhat less accurate solution than the stochastic numerical simulations of COMSALT.

For Mars condition, we find that the impact threshold is approximately an order of magnitude below the fluid threshold (Fig. 3b). This result is quantitatively consistent with the recent numerical result obtained with COMSALT [*Kok*, 2010], as well as qualitatively consistent with the studies of *Claudin and Andreotti* [2006] and *Almeida et al.* [2008].

To facilitate a simple calculation of the impact threshold in future studies, we approximate the result of our analytical model by the expression

$$u*_{it} = \left(\frac{700}{P}\right)^{\frac{1}{6}} \left(\frac{220}{T}\right)^{\frac{2}{5}} \exp\left(-5.1 + 280\sqrt{D_p} - 3.6 \cdot 10^3 D_p\right),$$  (27)

with a difference from the analytical result of ~1-10 % within the range $P = 500 – 1000$ Pa, $T = 180 – 270$ K, and $D_p = 100 – 1000$ μm.

### 4. Discussion and conclusions

The finding that the impact threshold is an order of magnitude below the fluid threshold has important implications for the formation of dunes, ripples, and dust storms, as discussed in more detail in *Kok* [2010]. Most importantly, it allows saltation transport to take place well below the fluid threshold, depending on whether the wind speed exceeded the fluid threshold more recently than that it dropped below the impact threshold. The occurrence of saltation transport at a given instantaneous wind speed intermediate between the impact and fluid thresholds is thus dependent on the history of the system, a phenomenon known as "hysteresis" [*Kok*, 2010]. It is critical to account for this hysteresis effect in calculations of Martian saltation, as it helps explain the various observations of active saltation on the Martian surface [*Kok*, 2010], despite the fact that Mars landers [*Zurek et al.*, 1992] and atmospheric simulations [*Fenton et al.*, 2005] indicate that wind speeds in excess of the fluid threshold occur very rarely on the Martian surface.

# REFERENCES

Almeida, M. P., et al. (2008), Giant saltation on Mars, Proc. Natl. Acad. Sci. U. S. A., 105(17), 6222–6226, doi:10.1073/pnas.0800202105.

Anderson, R. S., and P. K. Haff (1991), Wind modification and bed response during saltation of sand in air, Acta Mech., 1, 21–51

Atreya, S. K., et al. (2006), Oxidant enhancement in martian dust devils and storms: Implications for life and habitability, Astrobiology, 6(3), 439–450.

Bagnold, R. A. (1937), The transport of sand by wind, The Geographical Journal, 89(5), 409-438

Bagnold, R. A. (1941), The Physics of Blown Sand and Desert Dunes, Methuen, New York.

Cheng, N. S. (1997), Simplified settling velocity formula for sediment particle, *Journal of Hydraulic Engineering-Asce*, *123*(2), 149-152.

Claudin, P., and B. Andreotti (2006), A scaling law for aeolian dunes on Mars, Venus, Earth, and for subaqueous ripples, *Earth Planet. Sc. Lett., 252*(1-2), 30-44, doi:10.1016/j.epsl.2006.09.004.

Crane Company, Flow of Fluids through Valves, Fittings, and Pipe, **Technical Paper 410** (1988).

Creyssels, M., et al. (2009), Saltating particles in a turbulent boundary layer: Experiment and theory, J. Fluid Mech., 625, 47 – 74, doi:10.1017/S0022112008005491.

Fenton, L. K., A. D. Toigo, and M. I. Richardson (2005), Aeolian processes in Proctor Crater on Mars: Mesoscale modeling of dune-forming winds, J. Geophys. Res. 110, E06005.

Goudie, A.S., and N.J. Middleton (2006), *Desert dust in the global system*, Springer, Berlin.

Greeley, R., and J. D. Iversen (1985), Wind as a Geological Process on Earth, Mars, Venus, and Titan, Cambridge Univ. Press, New York.

Huang, N., et al. (2007), A model of the trajectories and midair collision probabilities of sand particles in a steady state saltation cloud, *Journal of Geophysical Research-Atmospheres*, *112*(D8), D08206.

Iversen, J. D., and K. R. Rasmussen (1994), The effect of surface slope on saltation threshold, Sedimentology, 41(4), 721–728, doi:10.1111/j.1365-3091.1994.tb01419.x.

Kok, J. F. (2010), Difference in wind speeds required for initiation versus continuation of sand transport on Mars: Implications for dunes and dust storms, *Phys. Rev. Lett.*, in press.

Kok, J. F., and D. J. Lacks (2009), Electrification of granular systems of identical insulators, *Physical Review E*, 79(5), 051304.

Kok, J. F., and N. O. Renno (2006), Enhancement of the emission of mineral dust aerosols by electric forces, Geophys. Res. Lett., 33, L19S10, doi:10.1029/2006GL026284.

Kok, J. F., and N. O. Renno (2008), Electrostatics in wind-blown sand, Phys. Rev. Lett., 100(1), 014501, doi:10.1103/PhysRevLett.100.014501.

Kok, J. F., and N. O. Renno (2009a), The electrification of wind-blown sand on Mars and its implications for atmospheric chemistry, Geophys. Res. Lett., 36, L05202, doi:10.1029/2008GL036691.

Kok, J. F., and N. O. Renno (2009b), A comprehensive numerical model of steady state saltation (COMSALT), J. Geophys. Res., 114, D17204, doi:10.1029/2009JD011702.

Nikuradse, J. (1933), Laws of flow in rough pipes (1950 translation), Tech. Memo. 1292, *Natl. Advis. Comm. on Aeronaut.*, Washington, D. C.

Prandtl, L. (1935), The mechanics of viscous flows, in *Aerodynamic Theory*, Vol. III, edited by W. F. Durand, Springer, Berlin.

Rasmussen, K. R., and M. Sorensen (2008), Vertical variation of particle speed and flux density in aeolian saltation: Measurement and modeling, J. Geophys. Res., 113, F02S12, doi:10.1029/2007JF000774.

Rice, M. A., et al. (1995), An experimental study of multiple grain-size ejecta produced by collisions of saltating grains with a flat bed, Sedimentology, 42(4), 695–706, doi:10.1111/j.1365-3091.1995.tb00401.x.

Ruf, C., et al. (2009), Emission of non-thermal microwave radiation by a Martian dust storm, Geophys. Res. Lett., 36, L13202, doi:10.1029/2009GL038715.

Shao, Y. P., and H. Lu (2000), A simple expression for wind erosion threshold friction velocity, *Journal of Geophysical Research-Atmospheres*, *105*(D17), 22437-22443.

Ungar, J. E., and P. K. Haff (1987), Steady-state saltation in air, *Sedimentology*, *34*(2), 289-299.

Wang, D. W., et al. (2008), Statistical analysis of sand grain/bed collision process recorded by high-speed digital camera, *Sedimentology*, *55*(2), 461-470.

White, B. R. (1979), Soil transport by winds on Mars, J. Geophys. Res., 84(B9), 4643–4651, doi:10.1029/JB084iB09p04643.

Zurek, R. *et al.* Dynamics of the atmosphere of Mars, in *Mars*, edited by H. H. Kieffer *et al.*, pp. 835-933, Univ. of Ariz. Press, Tucson, 1992.

**FIGURES**

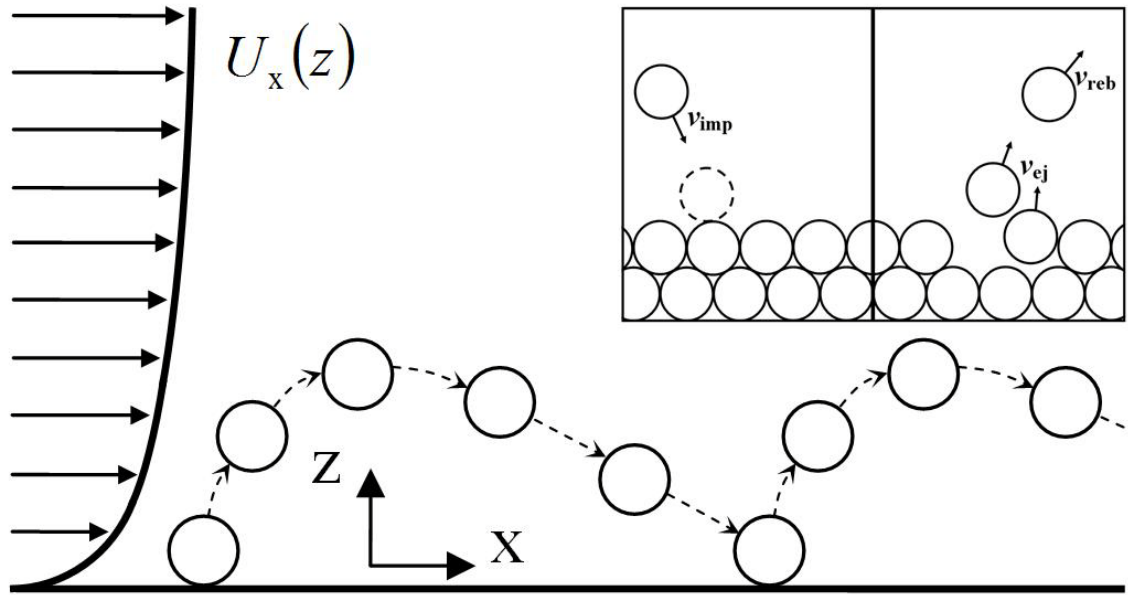

**Figure 1**. The idealized ballistic trajectory of a saltating sand particle propelled by the logarithmic wind profile $U_x(z)$ [e.g., *Kok and Renno*, 2009b] and bouncing along the soil surface. The inset shows a schematic representation of a saltating particle approaching the surface (left), rebounding from it, and ejecting (or splashing [*Ungar and Haff*, 1987]) several surface particles (right). After *Kok and Renno* [2009b].

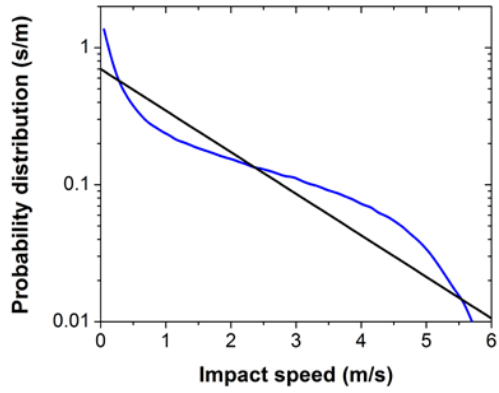

**Figure 2.** Simulations with the numerical saltation model COMSALT [*Kok and Renno*, 2009b] of the probability distribution of the impact speed of 250 μm saltating particles for Earth conditions at $u^* = 0.5$ m/s. The dashed black line denotes the approximate exponential distribution (Eq. (13)) with the simulated mean impact speed.

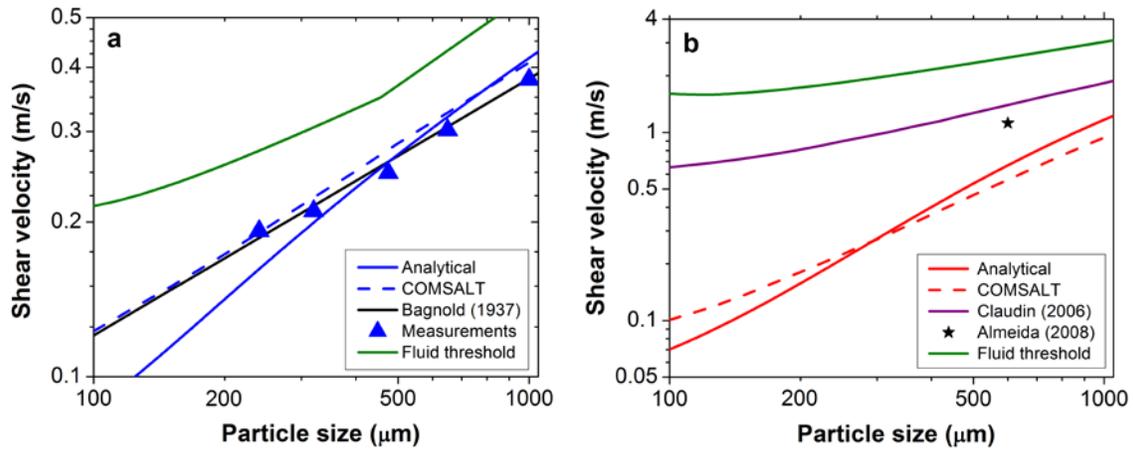

**Figure 3.** (a) Analytical calculation of the Earth impact threshold (solid blue line). Plotted for comparison are measurements [*Bagnold*, 1937; *Iversen and Rasmussen*, 1994; triangles], Bagnold's empirical relation [*Bagnold*, 1937; black line], the numerical simulation with COMSALT [*Kok and Renno*, 2009b; dashed blue line], as well as the fluid threshold [*Greeley and Iversen*, 1985; green line]. (b) Analytical calculation of the Mars impact threshold (solid red line). Plotted for comparison are numerical and analytical calculations of the impact threshold by *Kok* [2010; dashed red line], *Claudin and Andreotti* [2006; purple line], and *Almeida et al.* [2008; star], as well as the fluid threshold [*Greeley and Iversen*, 1985; green line]. Results are for an air pressure and temperature of 700 Pa and 220 K.